\documentclass[fleqn,twoside,twocolumn,nofootinbib,showkeys]{revtex4} 
\usepackage[nocpr]{ujp} 

\begin{document}
\title[Optical Properties of Irradiated Epitaxial GaN Films]
{OPTICAL PROPERTIES\\ OF IRRADIATED EPITAXIAL GaN FILMS}%
\author{A.E. Belyaev}
\affiliation{V.E. Lashkaryov Institute of Semiconductor Physics, Nat. Acad. of Sci. of Ukraine}
\address{41, Prosp. Nauky, Kyiv 03028, Ukraine}
\email{klyui@isp.kiev.ua}
\author{N.I.~Klyui}
\affiliation{V.E. Lashkaryov Institute of Semiconductor Physics, Nat. Acad. of Sci. of Ukraine}
\address{41, Prosp. Nauky, Kyiv 03028, Ukraine}
\email{klyui@isp.kiev.ua}
\author{R.V.~Konakova}
\affiliation{V.E. Lashkaryov Institute of Semiconductor Physics, Nat. Acad. of Sci. of Ukraine}
\address{41, Prosp. Nauky, Kyiv 03028, Ukraine}
\email{klyui@isp.kiev.ua}
\author{A.M.~Luk'yanov}
\affiliation{V.E. Lashkaryov Institute of Semiconductor Physics, Nat. Acad. of Sci. of Ukraine}
\address{41, Prosp. Nauky, Kyiv 03028, Ukraine}
\email{klyui@isp.kiev.ua}
\author{Yu.M.~Sveshnikov}
\affiliation{Close Corporation \textquotedblleft Elma-Malakhit\textquotedblright}
\address{Zelenograd, Russia}
\author{A.M.~Klyui\,}
\affiliation{Taras Shevchenko National University of Kyiv}
\address{2, Prosp. Academician Glushkov, Kyiv 03022, Ukraine}

\udk{538.958}  \razd{\secix}


\autorcol{A.E.\hspace*{0.7mm}Belyaev, M.I.\hspace*{0.7mm}Klyui,
R.V.\hspace*{0.7mm}Konakova et al.}

\setcounter{page}{34}%

\begin{abstract}
The influence of a microwave treatment (MWT) on the optical
properties of hexagonal GaN films has been studied. To estimate
the internal mechanical strains and the degree of structural
perfection in a thin near-surface layer of the film, the
electroreflectance (ER) method is used. The ER spectra are
measured in the interval of the first direct interband
transitions. It has been shown that the MWT results in the
relaxation of internal mechanical strains in the irradiated films.
In addition, the structural perfection in the thin near-surface
layer of the irradiated film became higher. A mechanism that
includes resonance effects and the local heating of the film
defect regions is proposed to explain the effects observed.
\end{abstract}

\maketitle

\section{Introduction}

Epitaxial films on the basis of A$^{\mathrm{III}}$B$^{\mathrm{V}}$
compounds~-- in particular, III-nitrides (GaN, AlN, BN) -- are
intensively used in modern micro-, nano-, and optoelectronics to
manufacture a wide class of devices \cite{1,2,3}. Gallium nitride
films are applied most successfully in the production of light and
laser diodes for the short-wave spectral range \cite{1}, as well as
in RF and high-temperature devices \cite{2,3}. In all cases, an
integral part of the technological process is the application of
different treatments to either the films themselves or the devices
in whole. It is evident that the research of the treatments effect
on the film properties is an important and challenging task, because
knowlenge of the mechanisms of film property modification allows the
parameters of the film-based devices to be
\mbox{predicted.}

Among the active treatments of GaN films, most widely used are the
thermal annealings (short- and long-term) \cite{4,5}, plasma,
chemical, and photochemical treatments \cite{6,7,8}.\,\,The
influence of penetrating radiation, e.g., $\gamma$-radiation, is
most often regarded not as a component of the fabrication
technology, but rather as a possible factor of their degradation. On
the other hand, as was shown in works \cite{9,10}, $\gamma
$-irradiation can improve the properties of epitaxial GaN films and
the structures on their basis.

The microwave treatment (MWT) is successfully applied to improve the
quality of GaN contacts \cite{11} and to carry out the
high-temperature activation of an acceptor impurity (e.g., Mg) in
GaN, which was introduced either in the course of growing \cite{12}
or at the subsequent ion implantation \cite{13}. In the opinion of
the author of work \cite{14}, MWT is the only method that allows
high-temperature treatments to be realized at extremely high rates
of temperature growth and decrease. However, this way is impossible
if the usual rapid thermal annealing is applied. In the latter case,
MWT equipments  based on powerful gyrotrons are used, which allows
high densities of radiation power to be obtained \cite{14}. At the
same time, it was demonstrated earlier that the low-power MWT can
also substantially affect the properties of silicon \cite{15} and
A$^{\mathrm{III}}$B$^{\mathrm{V}}$ films, as well as instrument
structures on their basis \cite{16,17}. Therefore, this work aimed
at studying the influence of the low-power MWT on the optical
properties of a thin near-surface layer in epitaxial GaN films.

\section{Experimental Part}

In our studies, we used GaN films with the $n$-type conductivity,
which were doped with silicon to concentrations of about
$(1\div3)\times10^{17}$~\textrm{cm}$^{-3}.$ The films were deposited
onto the (001) surface of Al$_{2}$O$_{3}$ substrates using the
metallorganic chemical vapor deposition (MOCVD) technique. The
thickness of the films was 1~$\mu\mathrm{m}$.

For MWT, we used UHF radiation with a frequency of 2.45~GHz. A magnetron
with a specific output power of 1.5~W/cm$^{2}$ served as a radiation source.
MWT of the films was carried out in the working chamber of a magnetron in the
air environment. The MWT cycles were carried out stage-by-stage with a gradual
increase of the irradiation time. The total time of MWT was equal to 5, 10, 20, and~30~s.

The electroreflection (ER) spectra were measured after every
radiation cycle. The measurements were carried out in the wavelength
range 350--380\textrm{~nm} (3.3--3.45~eV) and with a resolution of
$\pm0.001$~eV. The ER spectra were measured on an automated setup
based on a diffraction spectrometer MDR-23. An ohmic contact with
the GaN film was provided by sputtering a titanium film
100\textrm{~nm }in thickness followed by its annealing at a
temperature of 700$~^{\circ}\mathrm{C}$ and, then, sputtering a gold
film with thickness 100~nm. As a front contact, the KCl solution in
distilled water was used. An electric field was applied to the
specimen in the electrolytic cell with a platinum reference
electrode. In hexagonal GaN, there is the spin-orbit splitting of
the valence band near the absorption edge at point $\Gamma$ of the
Brillouin zone \cite{9,18,19}. Therefore, the ER signal is formed by
three main interband transitions usually designated as A, B, and C.
For the interpretation of spectra, we used the model of ER with
three direct transitions. The parameters of ER spectra were
determined by fitting the theoretical results to the experimental
data. This is a usual practice for the interpretation of
experimental results obtained by the modulation spectroscopy method
\cite{9,18,19}.

\begin{figure}%
\vskip1mm
\includegraphics[width=7cm]{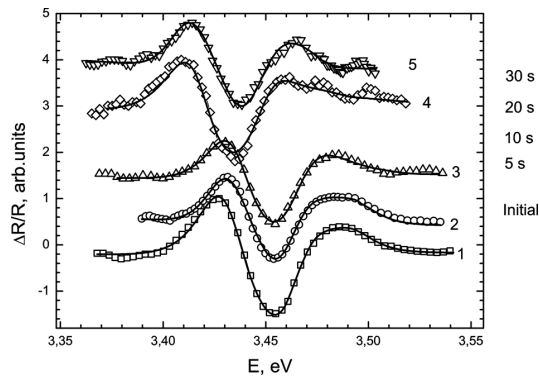}
\vskip-3mm\caption{ER spectra of GaN films before and after the
microwave treatment. Symbols denote experimental data, and the curve
exhibit the results of theoretical fitting: initial specimen
(\textit{1}) and after microwave irradiation for 5 (\textit{2}), 10
(\textit{3}), 20 (\textit{4}), and 30~s (\textit{5})  }
\end{figure}

\section{Results and Their Discussion}

It should be noted that, according to the results presented in work \cite{20}
and taking the conductivity of studied films into account, the penetration
depth of the microwave electromagnetic field exceeds the film thickness by several
orders of magnitude. Therefore, the power distribution across the film can be
regarded as uniform. A low temperature of a specimen during MWT can also
be taken as an important factor associated with the chosen MWT regimes.

The ER spectra of a GaN film before and after MWT are depicted in Fig.~1.
The symbols correspond to experimental data, and the solid curves to the
results of theoretical simulation (fitting). From Fig.~1, one can see that no
substantial changes are observed in the ER spectra at short MWT times; only
insignificant shifts of the peaks toward the short-wave region of the
spectrum take place (spectra~\textit{2} and \textit{3}). At the same time, the
increase of the treatment time to 20 and 30~s results in appreciable shifts of the
ER spectrum toward lower energies (spectra \textit{4} and \textit{5}).

These conclusions are directly confirmed by the energy dependences of
excitonic transitions A, B, and C obtained while fitting the theoretical
spectra to experimental ones (Fig.~2). One can see that, for MWT times of 5
and 10~s, the transition energies insignificantly grow, whereas the further
increase of the MWT time brings about a reduction of the transition energies to
values that are considerably lower than the initial ones. As was shown in work
\cite{9} on the basis of results obtained for a freestanding GaN film
\cite{18}, there are the squeezing stresses in the initial films. Hence, at the
first MWT stages, those stresses increase a little and, then, relax. A certain
growth of the internal mechanical stresses at short MWT times may be associated
with the generation of additional structural defects owing to the resonance
interaction between the UHF electromagnetic field and the defect subsystem of
the film, analogously to what was observed in works \cite{16,17}.

\begin{figure}%
\vskip1mm
\includegraphics[width=7cm]{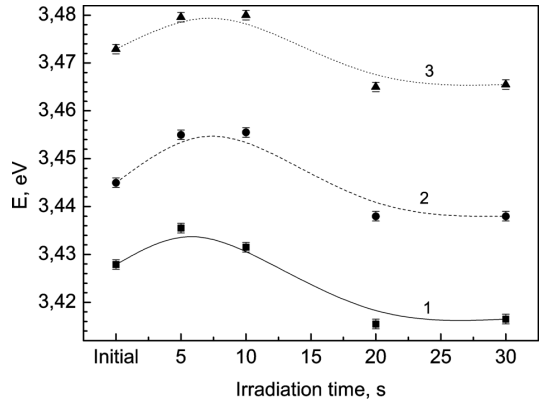}
\vskip-3mm\caption{Dependences of the direct transition energies on
the microwave treatment time. Curves \textit{1} to \textit{3}
correspond to interband transitions A, B, and C, respectively
}\vskip2mm
\end{figure}

\begin{figure}
\includegraphics[width=7cm]{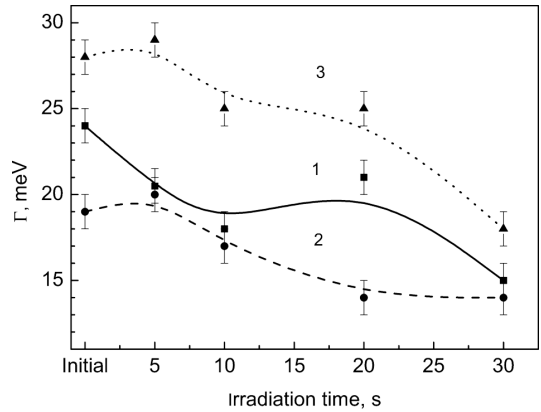}
\vskip-3mm\caption{The same as in Fig.~2, but for the
band-broadening parameter  }
\end{figure}

The increase of the MWT time gives rise to the warming up and the resonance activation
of local defect regions, when the MWT frequency coincides with the
characteristic frequencies of defect vibrations \cite{16,17}. Owing to a
considerable mismatch between the lattice constants in the film and the
substrate \cite{21}, there exists a gradient of mechanical stresses in the
system. As a result, the gettering of structural defects can take place at the
film--substrate interface. At a substantial stress relaxation degree, this
effect should disappear, which is also observed in the experiment. The
described mechanism is confirmed by the dependences of the band-broadening
parameter $\Gamma$ for the corresponding bands in the ER spectrum (Fig.~3).
Note that the parameter $\Gamma$ is directly related to the degree of
structural perfection in the film. The energy relaxation time for
light-excited charge carriers, $\tau$, (and, hence, their mobility) is related
to the band-broadening parameter as follows: $\tau\sim h/\Gamma$, where $h$
is Planck's constant \cite{9}. Therefore, the reduction of the transition
band-broadening, which is observed at long MWT times, confirms the mechanism
proposed above. The effects observed cannot be explained by purely thermal
processes. As was already mentioned above, the selected parameters of MWT
prevent the specimens from a considerable warming up, whereas rather high
temperatures are required for a rapid thermal annealing of defects in GaN
films to take place \cite{5}.

\section{Conclusions}

To summarize, a conclusion can be drawn that MWT, even the low-power one,
is a promising method for the improvement of the properties of epitaxial GaN
films, in particular, for a reduction of the mechanical stresses in them and an
enhancement of the structural perfection in the near-surface layer of a film.
The latter factor is especially important for the technology of device
production on the basis of III-nitrides. It is so because the GaN film is
often used as a substrate for the following epitaxial deposition of other
films, so that the properties of the surface and the near-surface region in
GaN substantially govern the characteristics of the instrument structure in
general \cite{1,2,3,21}.

\vspace*{-3mm}
\rezume{%
А.Є. Бєляєв, М.І. Клюй, Р.В. Конакова,\\ А.М. Лук'янов, Ю.М.
Свешніков, А.М. Клюй}{ОПТИЧНІ ВЛАСТИВОСТІ ЕПІТАКСІЙНИХ\\ ПЛІВОК GaN,
ЩО ПЕРЕБУВАЛИ ПІД ДІЄЮ\\ МІКРОХВИЛЬОВОГО ОПРОМІНЕННЯ} {Досліджено
вплив мікрохвильових обробок (МХО) на оптичні властивості
епітаксійних плівок GaN гексагональної модифікації. Для діагностики
рівня внутрішніх механічних напружень і структурної досконалості
тонкого приповерхневого шару плівки використовувався метод
електровідбивання (ЕВ). Спектри ЕВ вимірювалися в області
локалізації перших прямих зона-зонних переходів. Показано, що
внаслідок МХО в опромінених плівках спостерігається релаксація
внутрішніх механічних напружень і поліпшення структурної
досконалості приповерхневого шару. Запропоновано механізм виявлених
ефектів, що враховує резонансні ефекти і локальний розігрів
дефектних областей плівки.}

\end{document}